\documentclass[aps,pre,10pt,twocolumn]{revtex4-2}
\usepackage{graphicx}
\usepackage{epstopdf}
\usepackage{amsmath}
\usepackage{amssymb}
\usepackage[utf8]{inputenc}
\usepackage{graphicx,float,hyperref}
\usepackage{natbib}
\usepackage{xcolor}
\newcommand{\be}{\begin{equation}}
\newcommand{\bea}{\begin{eqnarray}}
\newcommand{\bc}{\begin{center}}            
\newcommand{\ee}{\end{equation}}
\newcommand{\eea}{\end{eqnarray}}
\newcommand{\ec}{\end{center}}

\newcommand{\baa}{\begin{eqnarray*}}
\newcommand{\eaa}{\end{eqnarray*}}
\begin{document}
\title{Effects of noise-induced coherence on the performance of a four-level laser heat engine}
\author{Kirandeep Kaur}
\email{kiran.nitj@gmail.com}
\affiliation{ Department of Physical Sciences, \\ 
Indian Institute of Science Education and Research Mohali,
Sector 81, S.A.S. Nagar, Manauli PO 140306, Punjab, India}

\begin{abstract}

In this work, we study the effect of noise-induced coherence on the performance analysis of a degenerate four-level  quantum heat engine with particular focus on the universal nature of efficiency, which refers to the appearance of the first two universal terms $\eta_C/2$ and $\eta_C^2/8$ in series expansion of the efficiency at maximum power under a left-right symmetry in the system.  Firstly, for a two-parameter optimization scheme, we derive an analytic expression for the efficiency at maximum power for the near-equilibrium condition and show that presence of noise-induced coherence breaks the left-right symmetry of the system. However, when the operation of the engine is restricted to either high-temperature or low-temperature regime, we discuss the conditions under which  the left-right symmetry can be retained in each case, giving rise to the universal characteristic of efficiency.  In case of one-parameter optimization, we show that while the universality of the first linear term $\eta_c/2$ is robust and holds consistently across all conditions, the universality of the quadratic term  $\eta_C^2/8$ depends on the constraints imposed on the control parameters.   Finally, we examine the behavior of power as a function of noise-induced coherence parameter highlighting the role of matter-field coupling in determining the suitable operation regime for the heat engine to reap the benefits of noise-induced coherence.

\end{abstract} 
\maketitle
\section{Introduction}

The concept of quantum heat engines was first articulated by Scovil and Dubois in 1959 when they showed the equivalence of a three-level maser to a heat engine. It was realized that when a three-level system (used as a working medium) is connected to two thermal baths at different temperatures, its operation is equivalent to a heat engine whose efficiency is bounded by Carnot efficiency \cite{Scovil1959}. It converts the incoherent thermal energy from heat reservoirs into a coherent laser output, harnessing the system's quantum properties for efficient energy transfer. This revolutionized the field of quantum heat engines/refrigerators and accelerated the study of various quantum physical systems, such as harmonic oscillator, particle in a box, two-level systems, and others, as potential working mediums \cite{Salamon2016,Rezek2006,Deffner2018,WangTLS2013,GuoWang2013,LongLiu2015,WangHe2013,Linke2005,Josefsson2018,Zhang2013,Lindenberg2010,Borga2012,George2018,OzgurRabi,OzgurSuperradiant,OzgurCorrelation}.

Despite its simplicity, it has remained a primary model for exploring  thermodynamics of quantum systems \cite{Geva1994,Geva1996,VsinghPRR,Kiran2021}. Recently, the model has been studied to understand the concept of synchronization in quantum systems \cite{JaseemSai2020}. Further, three-level quantum heat engine has been used as a prototype system to understand the violations of thermodynamic uncertainty relations \cite{SeifertTUR,Horowitz2020,Gingrich2016} 
in quantum systems \cite{VSinghTUR,Patrick2021,VuSaito2022,EuijoonLee2024}. This shows  the model’s continued relevance in advancing our understanding of quantum systems. 

The phenomenon of quantum coherence plays a fundamental role in distinguishing quantum behavior of a system from the classical behavior and enables the emergence of various counterintuitive phenomena such as superposition, interference, entanglement etc. Thus, maintaining the coherence is essential to exploit the quantum effects in cutting-edge technologies, such as quantum thermal machines, quantum cryptography and quantum computation \cite{OzgurQuantumFuel,Harbola2012,Petruccione2018,Ferdi2014pre,Segal2018,Scully2011,ScullyAgarwal2003, Kwon}. The noise-induced coherence is one such phenomenon where quantum interference effects, generated by environmental noise, lead to coherence between degenerate energy levels of a system. In the context of quantum heat engines (QHEs), it has been shown that the noise-induced coherence plays a pivotal role in enhancing performance metrics such as power output and efficiency. For instance, coherence plays a key role in enabling highly efficient energy transfer within photosynthetic systems\cite{Dorfmanscully}. Scully et al. established that breaking the detailed balance, through noise-induced coherence, can significantly increase the power output of laser/photocell quantum heat engines \cite{Scully2010,Scully2011,}. This has been validated through experimental studies of polymer solar cells \cite{Bittner_2014} as well as in nitrogen-vacancy-based microscopic quantum heat engines in diamonds\cite{NVcenter2019}. 

In this paper, we aim to examine the impact of noise-induced coherence on the efficiency characteristics of a four-level system with two exactly degenerate energy levels   and investigate the influence of this degeneracy and the resulting coherence on the efficiency of the system. In particular, we focus on the universal nature of efficiency \cite{Esposito2009}, a common feature in both classical and quantum heat engines. Although the analytic expression for the efficiency at maximum power (EMP) is different in different models of heat engines, the first two terms of the Taylor series expansion of efficiency are identical, i.e.,  $\eta^{\rm MP}=\eta_C/2+\eta_C^2/8+O(\eta_C^3)$ \cite{Esposito2009,VsinghPRR,CA1975,Esposito,Esposito2010,VarinderJohal2018,Tu2008,Schmiedl2008}. This behavior of efficiency is known as the universal nature of efficiency \cite{Esposito2009, VandenBroeck2005, Hernandez2001,VsinghPRR}. 
While the tight-coupling condition (no heat leaks) in heat engines is sufficient to warrant the appearance of the first universal term $\eta_C/2$ \cite{Broeck2005}, the appearance of second universal term $\eta_C^2/8$  can be established in the presence of a certain kind of left-right symmetry in the system \cite{Esposito2009}.

The paper is organized as follows. In sec. II, we introduce the model of four-level maser heat engine. Sec. III is devoted to the study of effect of noise-induced coherence on the universal nature of efficiency. In Sec. IV, we discuss the optimization of power of the heat engine with respect to the noise-induced coherence parameter $p$. We conclude in Sec. V.

\section{Model of Four-Level Laser Quantum Heat Engine}
\begin{figure}   
 \begin{center}
\includegraphics[width=8.6cm]{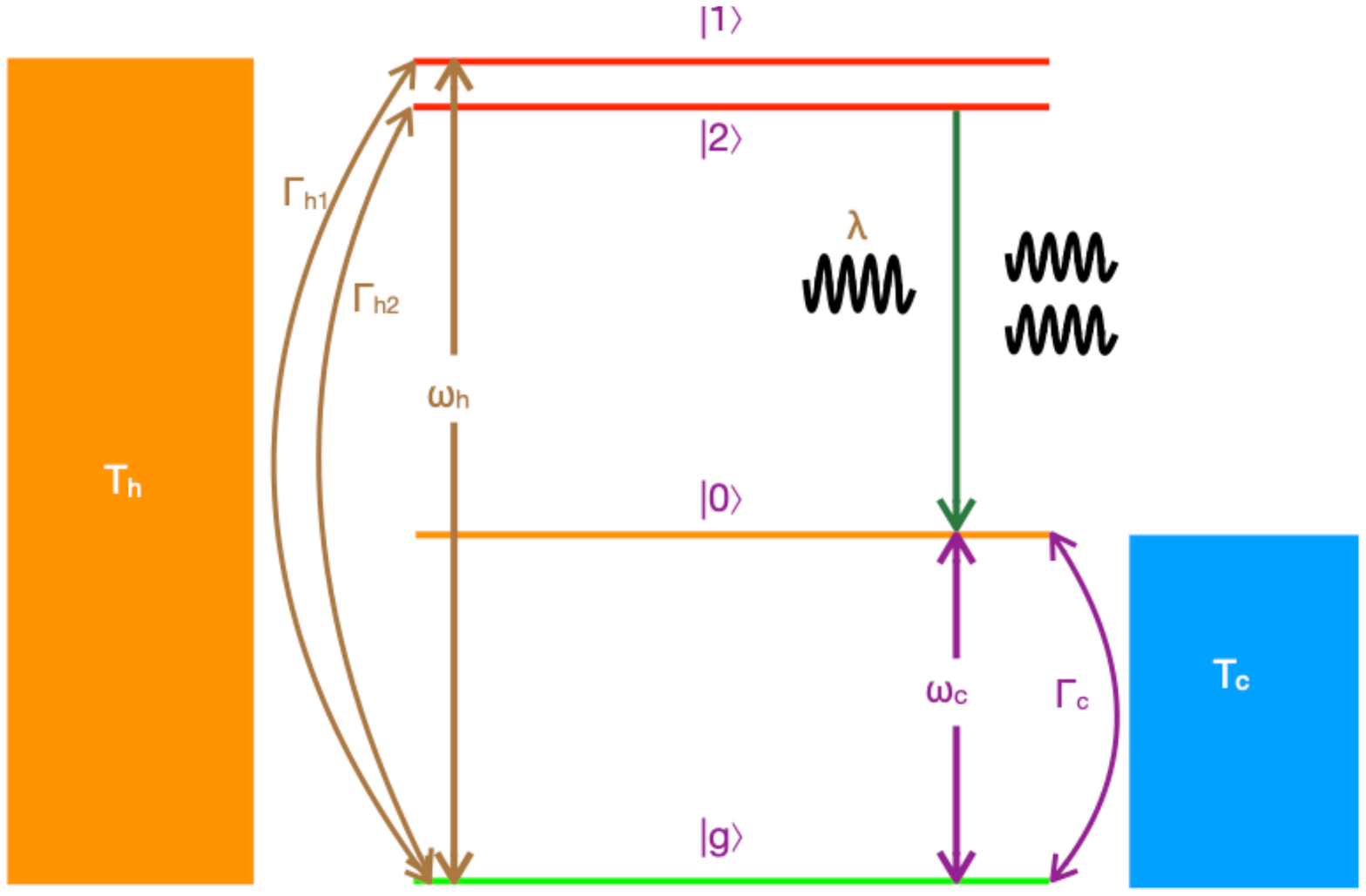}
 \end{center}
\caption{(Color online) Model of four-level laser heat engine with two-degenerate energy levels ($\vert 1\rangle$ and $\vert2\rangle$) continuously coupled to two heat reservoirs at temperatures $T_h$ and $T_c$.}
\end{figure}

This model consists of a four-level quantum system with two degenerate energy levels, two thermal reservoirs at temperature $T_c$ and $T_h$ such that $T_c < T_h$, and a semi-classical field  having frequency $\omega = \omega_2 - \omega_0$, see Fig. 1. Here, ground state, $\vert g\rangle$ and the degenerate upper levels $\vert 1\rangle$ and 
 $\vert 2\rangle$ are in thermal contact with the hot reservoir through a frequency filter that passes frequency $\omega_h = \omega_2 - \omega_g = \omega_1 - \omega_g$. The energy levels $\vert 0\rangle$ and  $\vert g\rangle$  are in thermal contact with cold reservoir through a frequency filter that passes frequency $\omega_c = \omega_0 - \omega_g$. $\omega_k$ is the atomic frequency of each state where $k=g,0,1,2$. The hot reservoir at temperature $T_h$ induces transitions between the ground state and degenerate energy states ($\vert g\rangle\leftrightarrow\vert1\rangle$ and $\vert g\rangle\leftrightarrow\vert2\rangle$), while the cold reservoir at temperature $T_c$ de-excites the intermediate level to ground state ($\vert 0\rangle\leftrightarrow\vert g\rangle$). Power generation in the system is driven by the interaction mediated by the classical field continuously coupled between the intermediate state ($\vert 0\rangle$) and degenerate states ($\vert 1\rangle$ and $\vert2\rangle$). The form of the interaction Hamiltonian is given by $V(t)=\hbar \lambda (e^{-i\omega t} \left(\vert 1\rangle\langle 0\vert + \vert 2\rangle\langle 0\vert \right) + \rm {h.c.})$, where $\lambda$ is matter-field coupling constant and  h.c. stands  for the hermitian conjugate.
 
 In a rotating frame  with respect to the bare Hamiltonian $H_0$, the dynamics of the engine is governed by following Lindblad master equation \cite{Lindblad1976,GKS76}:
\begin{equation}
\dot{\rho_R} = -\frac{i}{\hbar} [V_R,\rho] + \mathcal{L}_{h}[\rho] + \mathcal{L}_{c}[\rho],
\end{equation}
where $V_R=\hbar\lambda \left(\vert 1\rangle\langle 0\vert + \vert 2\rangle\langle 0\vert  + \vert 0\rangle\langle 1\vert + \vert 0\rangle\langle 2\vert\right) $, $\mathcal{L}_{c}$ and $\mathcal{L}_{h}$ are the Lindblad superoperators describing   the dissipative interaction with the hot and cold reservoirs, respectively:
\begin{widetext}
\begin{eqnarray}
\mathcal{L}_c[\rho]= \Gamma_c(n_c+1) \Big(A_c\rho A_c^\dagger  -\frac{1}{2}\big\{ A_c^\dagger A_k,\rho \big\}\Big) 
+  \Gamma_c n_c\Big(A_c^\dagger \rho A_c  -\frac{1}{2}\big\{ A_c A_c^\dagger,\rho \big\}\Big)  , \label{disscold}
\end{eqnarray}
\begin{eqnarray}
\mathcal{L}_h[\rho]=\sum_{k={1,2}}\Gamma_{hk}\Big[(n_h+1)\Big(A_k\rho A_k^\dagger  -\frac{1}{2}\big\{ A_k^\dagger A_k,\rho \big\}\Big) 
+  n_h\Big(A_k^\dagger \rho A_k  -\frac{1}{2}\big\{ A_k A_k^\dagger,\rho \big\}\Big)\Big] \nonumber
\\
+  \Gamma\cos\theta \Big[(n_h+1)\Big(A_1\rho A_2^\dagger -\frac{1}{2}\big\{ A_2^\dagger A_1,\rho \big\}\Big) 
 +
n_h\Big(A_1^\dagger \rho A_2  -\frac{1}{2}\big\{ A_2 A_1^\dagger,\rho \big\}\Big)\Big] \nonumber
\\
+  \Gamma\cos\theta \Big[(n_h+1)\Big(A_2\rho A_1^\dagger -\frac{1}{2}\big\{ A_1^\dagger A_2,\rho \big\}\Big) 
 +
n_h\Big(A_2^\dagger \rho A_1  -\frac{1}{2}\big\{ A_1 A_2^\dagger,\rho \big\}\Big)\Big] \label{disshot}
\end{eqnarray} 
\end{widetext}
where  $A_c=\vert g\rangle \langle 0\vert$,  $A_k=\vert g\rangle \langle k \vert$ ($k=1, 2$) are jump operators that govern the relevant transitions between the energy levels; $\Gamma_{h1(2)}$ and $\Gamma_c$ are system-bath coupling constants \cite{VsinghPRR}. Here, $\cos{\theta}$ represents the angle between
dipole transitions $\vert 1\rangle\rightarrow{\vert g\rangle}$ and $\vert 2\rangle\rightarrow{\vert g\rangle}$. The mean number of photons in the hot (cold) bath  corresponding to mode frequency $\omega_{h}$ ($\omega_c$) follows Planck's distribution given by $n_{h, c}=(e^{\hbar\omega_{h,c}/k_B T_{h,c}}-1)^{-1}$.

In this model,  output power and  the heat flux can   be defined by using the formalism develepoded by Tannor and Boukobza \cite{BoukobzaTannor2006A,BoukobzaTannor2006B,BoukobzaTannor2007}
\begin{eqnarray}
P &=& \frac{i}{\hbar} {\rm Tr}([H_0,V_R]\rho_R), \label{power1} \\
\dot{Q_h} &=&  {\rm Tr}(\mathcal{L}_h[\rho_R]H_0), \label{heat1} \\
\eta &=& \frac{P}{\dot{Q_h}}, \label{efficiency0}
\end{eqnarray}
where we have followed the sign convention in which heat entering the system, heat rejected by the system and the work extracted from the engine is taken to be positive. Evaluating the traces  appearing in Eqs. (\ref{power1}) and (\ref{heat1}), the power and heat flux take the following form:
\begin{eqnarray}
P &=&i\hbar\lambda(\omega_h-\omega_c)[(\rho_{01}-\rho_{10})+(\rho_{02}-\rho_{20})], \label{power222} \\
\dot{Q_h} &=&  i\hbar\lambda\omega_h [(\rho_{01}-\rho_{10})+(\rho_{02}-\rho_{20})],\label{heat2}
\end{eqnarray}
where $\rho_{0k} = \langle 0\vert\rho_R\vert k\rangle$ and $\rho_{k0} = \langle k\vert \rho_R\vert 0\rangle$ ($k=1, 2$). Finally, the expression for the efficiency is given by
\begin{equation}
\eta = \frac{P}{\dot{Q}_h} =1 - \frac{\omega_c}{\omega_h}. \label{efficiency}
\end{equation}
From Eq.(\ref{powerfinal}), the positive power production condition implies that $\omega_c/\omega_h\geq T_c/T_h$.
Hence $\eta \leq \eta_{\rm C}^{}$.
\section{Universal nature of the efficiency}

In this section, we explore the conditions leading to the universal nature of efficiency in the presence of noise-induced coherence in the engine. The expression for power output of the engine has been derived in Appendix A. Maximization of the power with respect to control variables $\omega_c$ and $\omega_h$ results in considerably complex equations, which cannot be solved to give analytical solution. However, for near-equilibrium condition, an analytic expression for the efficiency at maximum power can be obtained upto second order term in $\eta_C$, which is good enough for our purpose. 

First, we outline the formalism  developed by Esposito and co-authors, who proved the existence of first two universal terms in the Taylor expansion of the EMP  for the tight-coupling heat engines having a left-right symmetry in the system.  Following is the brief outline of the procedure developed in Ref. \cite{Esposito2009}. 

Let $I_E (x,y)$  and $I(x,y)$ are the average energy and average matter fluxes entering the system from the reservoir., where $x$ and $y$ are dimensionless scaled energies. If energy is transported explicitly by the particles (photons here) of given energy $\epsilon$, then we can write
\begin{equation}
    I_E (x,y) = \epsilon I(x,y). \label{fluxmatter}
\end{equation}
In this case, the general formula for the EMP, derived in Ref. \cite{Esposito2009}, is given by
\begin{equation}
\eta = \frac{\eta_C}{2} + \left( 1 + \frac{M}{\partial_{x}L} \right)\frac{\eta_C^2}{4} + \mathcal{O}(\eta_C^3),
\label{effas}
\end{equation}
where
$L=-I'_{1}(x,x)$ and $M=I''_{11}(x,x)/2$. $I(x,y)$. The prime superscript denotes the order of the derivative and the subscript indicates that the derivative is taken with respect to the first or second variable. When we impose the left-right symmetry on the system, the inversion of flux takes place, resulting in
\begin{equation}
I(x,y) = -I(y,x),   \label{fluxreversal}
\end{equation}
which further establish the condition $2M=-\partial_x L$, and the second term in Eq. (\ref{effas}) is reduced to $\eta_C^2/8$. This establishes the universality of the term $\eta_C^2/8$ under the symmetry described by Eq. (\ref{fluxreversal}).

We use the above formalism to explore the universality of efficiency of a degenerate four-level  maser heat engine in the presence of noise-induced coherence.  The first step towards our goal is to identify the flux term for the engine under consideration. In our model, flux of photons carry the energy flux from hot to cold reservoir.   Comparing Eq. (\ref{powerfinal})) with Eq. (\ref{fluxmatter}), we identify the flux term, $I$ as follows
\begin{widetext}
\begin{equation}
I=\frac{8\Gamma_c\Gamma_h\lambda^2 (p+1) e^y \left(e^y-1\right) \left(e^x-e^y\right) \left(e^x-1\right) } {A'+B'} , \label{fluxx1}
\end{equation}
where 
\begin{eqnarray}
 A' &=&   e^y \left(e^{x+y}+2 e^x+e^y\right) \left[e^x \left(e^y-1\right) \Gamma _c+(p+1) e^y\left(e^x-1\right)  \Gamma _h\right] (1+p) \Gamma _c \Gamma _h, \nonumber
 \\
 B' &=& 8 \lambda ^2 \left(e^x-1\right) \left(e^y-1\right) \left\{\left(e^y-1\right) \left[\left(e^x+2\right) e^y+e^x\right]\Gamma _c +(p+1) e^y \left(e^x-1\right)  \left(e^y+3\right) \Gamma _h\right\}.
\nonumber
\end{eqnarray}
Here,   $x\equiv \hbar\omega_c/k_B T_c$, $y\equiv \hbar\omega_h/k_B T_h$,  $n_h=1/(e^y-1)$ and $n_c=1/(e^x-1)$ have been used.  For the current $I$ given in Eq. (\ref{fluxx1}),   expressions for $L$, $M$, and 
$\partial_x L$ can be obtained, and substituting them in Eq. (\ref{effas}), we arrive at
 \begin{eqnarray}
 \eta^{\rm MP} &=&  \frac{\eta_C}{2} + \Bigg\{\frac{ e^{2 \alpha}\left[\left(e^\alpha-1\right)^2 \Gamma _c+(p+1) \left(e^\alpha-5\right) \left(e^\alpha+1\right) \Gamma _h\right](p+1)  \Gamma _c \Gamma _h}{\left(\Gamma _c+(p+1) \Gamma _h\right) \left[(p+1) e^{2 \alpha} \left(e^\alpha-3\right) \left(e^\alpha+1\right) \Gamma _c \Gamma _h+8 \lambda ^2 \left(e^\alpha-1\right)^2 \left(e^{2 \alpha}+3\right)\right]} \nonumber
 \\
 && + \frac{8 \lambda ^2 \left(e^\alpha-1\right)^2 \left(\left(e^\alpha \left(e^\alpha+2\right)+5\right) \Gamma _c+(p+1) \left(e^\alpha-3\right) \left(e^\alpha-1\right) \Gamma _h\right)}{ \left(\Gamma _c+(p+1) \Gamma _h\right) \left[(p+1) e^{2 \alpha} \left(e^\alpha-3\right) \left(e^\alpha+1\right) \Gamma _c \Gamma _h+8 \lambda ^2 \left(e^\alpha-1\right)^2 \left(e^{2 \alpha}+3\right)\right]}\Bigg\}\frac{\eta_C^2}{8}
 + O(\eta_C^3),
 \label{uni1}
  \end{eqnarray}
 \end{widetext}
 where $\alpha$ is the solution of the transcendental equation obtained by putting $L$ and $\partial_x L$ in equation $x=-2L/\partial_x L$. Now we will check the conditions under which the symmetry criterion in Eq. (\ref{fluxreversal}) is fulfilled for the model under consideration. However, before proceeding further, we would like to mention that for the non-degenerate, three-level maser heat engine, the symmetry condition (reversal of flux) in Eq. (\ref{fluxreversal}) is satisfied for $\Gamma_h=\Gamma_c$. On the contrary, by inspecting Eq. (\ref{fluxx1}), we can confirm that the condition $\Gamma_c=\Gamma_h$ is not sufficient to ensure the reversal of flux, implying that the presence of noise-induced coherence  breaks the left-right symmetry in the system. Interestingly, even for $p=0$, $I(x, y)\neq -I(y, x)$.  Further, inspecting Eq. (\ref{uni1}) for $\Gamma_h=\Gamma_c$ and $p=0$, we can confirm that the term inside the square brackets is not equal to unity. This implies that the even when noise-induced coherence parameter is set to $p=0$, the second universal term $\eta_C^2/8$ cannot be retained. In the following, we will explore the conditions for which the symmetric condition given in Eq. (\ref{fluxreversal}) is satisfied. 

 \subsection{High-temperature limit}
Here, we show that in addition to the conditions $\Gamma_h=\Gamma_c$ and $p=0$, if we restrict our engine to operate in the high-temperate regime, the symmetry condition  ($I(x, y) = -I(y, x)$) is satisfied and the second universal term $\eta_C^2/8$ appears in the series expansion of the EMP. In the high-temperature regime, we can set  $e^x=1+x$ and $e^y=1+y$ and the expression for flux (Eq. (\ref{fluxx1})) takes the following form:
 \begin{equation}
 I=\frac{  (p+1) x y \Gamma _c \Gamma _h (x-y)\lambda^2}{2 \left(y \Gamma _c+(p+1) x \Gamma _h\right) \left((p+1) \Gamma _c \Gamma _h+2 \lambda ^2 x y\right)}. \label{fluxx2}
  \end{equation}
 By inspecting Eq. (\ref{fluxx2}), it can be explicitly seen that the symmetry criterion in Eq. (\ref{fluxreversal}) is fulfilled for the $\Gamma_c=\Gamma_h$ and $p=0$. Hence, we expect second universal term $\eta_C^2/8$ under the abovesaid conditions. Let us look at the expression for the EMP for $\Gamma_h=\Gamma_c$, which can be derived by following the same steps  used for deriving Eq. (\ref{uni1}:  
 \begin{equation}
 \eta^{\rm MP}_{\rm HT} =   \frac{\eta_C}{2} + \frac{\eta_C^2(1+p)}{4(2+p)} + O(\eta_C^3). \label{uni2}
 \end{equation}
 Above equation clearly shows that the second universal term $\eta_c^2/8$ can be retained in the absence of the noise-induced coherence ($p=0$) , whose presence ($p\neq0$) is responsible for breaking the left-right symmetry in the system.

\subsection{Low Temperature Regime}
Although it is not possible to obtain an exact analytic solution for the two-parameter optimization scheme (optimization with $\omega_h$ and $\omega_c$) in general conditions, a closed form expression for the EMP can be obtained if we restrict ourselves to the low-temperature regime. In this case, we do not have to resort ourselves to the perturbative method used in the last section as an exact solution is available.   In the low-temperature limit, the form of power is given by
\begin{equation}
P_{LT} = F\, T_h[y - x (1-\eta_C)]\left( e^{-y}-e^{-x} ,  \right)
\label{ecocold}
\end{equation}
where constant $F\equiv F(\Gamma_c, \Gamma_h, \lambda, p)$
where  $F$ is the multiplicative constant having the functional form  $F=F(\Gamma_c, \Gamma_h, \lambda, p)$.

From Eq. ({\ref{ecocold}), the matter flux term can be identified as
\begin{equation}
I_{LT}(x,y) =  e^{-y}-e^{-x} . \label{LTsymmetry}
\end{equation}
Exchanging $x$ and $y$ in Eq. (\ref{LTsymmetry}) will lead to inversion of the flux, $I(x,y)=-I(y,x)$, thus we  expect the presence of the first two universal terms in the Taylor expansion of the EMP. This can be verified as follows. Optimization  of Eq. (\ref{ecocold}) with respect to $x$ and $y$ yields the following expressions for $x$ and $y$, respectively
\begin{equation}
    x = \frac{\eta_C-\ln(1-\eta_C)}{\eta_C}, \,\, y = \frac{\eta_C-(1-\eta_C)\ln(1-\eta_C)}{\eta_C}. \label{solxy}
\end{equation}
Substituting $x$ and $y$  in $\eta=1-(1-\eta_C)x/y$, the expression for the EMP can be obtained as
\begin{equation}
\eta^{\rm MP}_{\rm LT} = \frac{\eta_C^2}{\eta_C - (1-\eta_C)\ln{(1-\eta_C)}} 
\approx 
\frac{\eta_C}{2} + \frac{\eta_C^2}{8} + \frac{7\eta_C^3}{96} + ..,
\end{equation}
which contains the first two universal terms $\eta_C/2$ and $\eta_C^2/8$. 

We conclude this subsection by commenting that the   left-right symmetry in the system can be retained both in the high- and low-temperature regime for our degenerate four-level maser heat engine.  
\subsection{Universality of efficiency in one parameter optimization scheme}
In Sec. III A and III B, we have investigated the universality of the efficiency for the two-parameter optimization scheme. However, sometimes it is easy to obtain analytic solution for optimization with respect to one parameter only while keeping the other one fixed at a constant value. In this section, we explore the existence of universal nature of efficiency in the presence of noise-induced coherence for one-parameter optimization scheme.  In order to obtain analytic results, we assume that the system is governed by strong matter-field coupling where $\lambda \gg \Gamma_h,c$. Additionally, we restrict ourselves to the high-temperature regime where $n_h \approx k_B T_h/ \hbar\omega_h\gg 1$ and $n_c \approx k_B T_c/ \hbar\omega_c\gg 1$ .  Using Eq. (\ref{powerfinal}), the expression for power output in the abovesaid limit is obtained as
    \begin{equation}
   P =  \frac{(\omega_h-\omega_c)(\omega_c-(1-\eta_C)\omega_h)(1+p) \Gamma_h\Gamma_c}{4(1-\eta_C)\Gamma_c\omega_h+4(1+p)\Gamma_h\omega_c}. \label{power2} 
\end{equation} 
Again writing in terms of $y=\omega_h/T_h$ and $x=\omega_c/T_c$, the flux term can be identified as
\begin{equation}
    I (x, y) =  \frac{(x-y)(1+p)\lambda^2 \Gamma_h\Gamma_c}{4 y \Gamma_c  + 4 x \Gamma_h (1+p)}. \label{fluxht}
\end{equation}
Inspection of Eq. (\ref{fluxht}) reveals that for $\Gamma_h=\Gamma_c$ and $p=0$, switching of the variables $x$ and $y$ leads to the symmetric condition (or reversal of flux) , i. e., $I(x, y)=    -I(y, x)$. In the following,  we show that, unlike the two-parameter optimization scheme,  reversal of flux does not ensure the universal nature of efficiency for one-parameter optimization scheme.

Having said this, now be turn to study performance of the engine for one-parameter optimization scheme. Either we can fix $\omega_h$ or $\omega_c$. First, for fixed value of $\omega_h$, optimization of Eq. (\ref{power2}) with respect to $\omega_c$ yields the following expression for the EMP:
  \begin{widetext}
    \begin{equation}
 \eta^{\rm MP}_{\omega_h}= \frac{\Gamma_h(1+p)+\Gamma_c(1-\eta_C)-\sqrt{(1-\eta_C)[\Gamma_c+\Gamma_h(1+p)][(1-\eta_C)\Gamma_c+(1+p)\Gamma_h]}}{(1+p)\Gamma_h}.
\label{effwh}
\end{equation}
Here, the subscript $\omega_h$  denotes that it is fixed. We will follow this convention throughout the text. To explore the universality of efficiency, we expand Eq. (\ref{effwh}) in Taylor's series:
\begin{equation}
   \eta^*_{\omega_h=k} = \frac{\eta_C}{2} + \frac{\eta_C^2(1+p)\Gamma_h}{8[\Gamma_c+\Gamma_h(1+p)]}  +  \frac{\eta_C^3(1+p)\Gamma_h[2\Gamma_c+\Gamma_h(1+p)]}{16[\Gamma_c+\Gamma_h(1+p)]} + O(\eta_C^4) \label{taylor1}.
\end{equation}
Similarly,   optimization of Eq. (\ref{power2}) with respect to $\omega_h$ for a fixed $\omega_c$ results in the following series expansion,
\begin{equation}
  \eta^{\rm MP}_{\omega_c}=    \frac{\eta _C}{2}+\frac{\eta _C^2 \left(\Gamma _c+2 (p+1) \Gamma _h\right)}{8 \left(\Gamma _c+\Gamma _h+p \Gamma _h\right)}+\frac{\eta _C^3 \left(\Gamma _c^2+4 (p+1) \Gamma _c \Gamma _h+2 (p+1)^2 \Gamma _h^2\right)}{16 \left(\Gamma _c+\Gamma _h+p \Gamma _h\right){}^2}+O\left(\eta _C^4\right) \label{taylor2}.
\end{equation}
We notice that in both the above equations,  we retain the first universal term $\eta_C/2$. The presence of first universal term $\eta_c/2$ is natural as our system obeys tight-coupling (no heat leaks) condition. However, even imposing the the condition $\Gamma_c=\Gamma_c$ along with $p=0$ (which ensures the reversal of flux given in Eq. (\ref{fluxht})), the second terms in Eqs. (\ref{taylor1}) and (\ref{taylor2}) respectively reduce to $\eta_C^2/16$ and $3\eta_C^2/16$, which is different from the second universal term $\eta_C^2/8$.  Thus, unlike the two-parameter optimization schemes discussed earlier, the one parameter optimization scheme does not result in the second universal term for $\Gamma_c=\Gamma_h$ and $p=0$.

In the preceding discussion, we demonstrate that incorporating an extra symmetric constraint \cite{Uzdin_universal2014, VsinghPRR, Parkuniversal} on the parametric space along with the conditions $\Gamma_c=\Gamma_h$ and $p=0$ leads to the emergence of the second universal term. To achieve this, we consider two types of symmetric constraints, one is $\omega_c + \omega_h = k$ and another is $\omega_c\omega_h = k$.  For the case $\omega_h+\omega_c= k$, the one parametric optimization scheme yields
\begin{equation}
    \eta^*_{\omega_h+\omega_c} = \frac{\eta_C}{2}+\frac{\eta _C^2 \left(\Gamma _c+3 (p+1) \Gamma _h\right)}{16 \left(\Gamma _c+\Gamma _h+p \Gamma _h\right)}+\frac{\eta _C^3 \left(\Gamma _c^2+10 (p+1) \Gamma _c \Gamma _h+5 (p+1)^2 \Gamma _h^2\right)}{64 \left(\Gamma _c+\Gamma _h+p \Gamma _h\right){}^2} +.... \label{taylor3}
\end{equation}
 \end{widetext}
Substituting $p=0$ and $\Gamma_h=\Gamma_c$ in Eq. (\ref{taylor3}), the second term reduces to $\eta_C^2/8$, thus confirming the universal nature of efficiency. Similarly, we can confirm the appearance of the second universal term for the symmetric constrain $\omega_h\omega_c=k$ as well. This suggests that the left-right symmetry should not be limited to the reversal of flux only, but it may also be retained by imposing symmetric constraints on the parameter space.

\section{Optimization with respect to noise-induced coherence parameter $p$}
In the previous works \cite{Scully2011,Dorfman2018}, the performance of the SSD engine is studied in the high-temperature and strong coupling regime ($\lambda>>\Gamma_{h,c}$). In the abovesaid regime, the power output of the heat engine is monotonically increasing function of the noise-induced coherence parameter $p$. Here, we relax these assumptions and study the performance of the SSD engine under the general conditions. The expression for the power output of the engine is given by [see Appendix A]
\begin{equation}
P = \frac{4(n_h - n_c)(1+n_h)(1+p)\Gamma_c \Gamma_h\lambda^2 (\omega_h -\omega_c)}{T_1 + T_2 +T_3 + T_4}, \label{pp1}
\end{equation}
where $T_1 = 2\lambda^2\Gamma_h(1+n_h)(1+4n_h)(1+p)$, 
$T_2 = 2\lambda^2\Gamma_c(1 + 3n_c + 2 n_h + 4n_c n_h)$, 
$T_3 =\Gamma_c^2 \Gamma_h (1+n_c)(1+n_h)(1+3 n_h + n_c(2+4n_h))(1+p)$ and
$T_4 = \Gamma_c\Gamma_h^2(1+n_h)^2 (1+2n_c + 3n_h + 4n_c n_h) (1+p)^2$. 
Optimization of the power of power $P$ with respect to $p$ ($\partial P/\partial p=0$) yields the following optimal expression for the parameter $p$:
\begin{equation}
p^* = \sqrt{\frac{2(1+3n_c + 2n_h + 4n_c n_h)}{1+3n_h+2n_c + 4n_c n_h}} \frac{\lambda}{\Gamma_h(1+n_h)} -1.  \label{optimalp}
\end{equation}
We graphically illustrate our point by plotting  power output as a function of noise-induced coherence parameter $p$ in Fig. 2 by fixing all other parameters. Further in high temperature limit ($n_c,n_h >> 1$) and low temperature ($n_c,n_h << 1$) limit, Eq.(\ref{optimalp}) takes the following forms, respectively:
\begin{equation}
p^*_{HT} = \frac{\sqrt{2}\lambda}{\Gamma_h n_h}-1, 
\qquad
p^*_{LT} = \frac{\sqrt{2}\lambda}{\Gamma_h}- 1. 
\end{equation}
In the strong-coupling regime ($\lambda\gg \Gamma_h$), $p^*_{LT}>1$, which implies that for the physical range of $p$, $-1\leq\,p\,\leq\,1$, the power output will be monotonically increasing function of the  the noise-induced coherence parameter $p$. Same conclusion will also hold true for the high-temperature regime provided $\lambda\gg \Gamma_h\, n_h$.  This implies that in the strong coupling regime, engine should be operated close to $p=1$. This can be clearly seen from Fig. 2 that for increasing fixed value of $\lambda$, $\lambda=0.1$ (solid gray curve), $\lambda=0.2$ (dashed blue curve) $\lambda=0.3$ (dotted red curve), the maximum   power point shifts to the right towards higher value of $p$. 

However, for weak coupling regime ($\lambda\ll \Gamma_h$),  $p^*_{HT(LT)} < -1$.   Hence, within physical range of p (-1,1), the power will be monotonically decreasing function of $p$. Thus, contrary to the strong-coupling regime, engine should be operated close to $p=-1$ to reap the benefits of noise-induced coherence in increasing the power output. 
\begin{figure}   
 \begin{center}
\includegraphics[width=8.6cm]{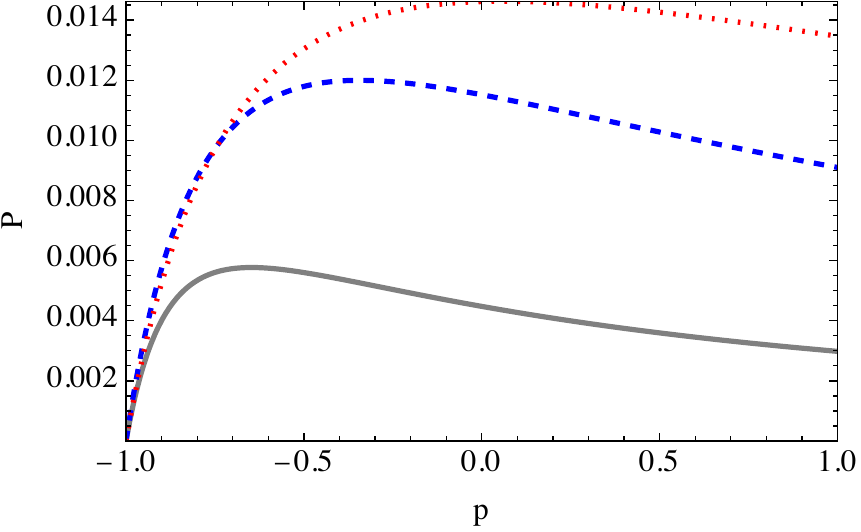}
 \end{center}
\caption{Power as a function of noise-induced coherence parameter $p$. Here, $\Gamma_c=0.25$, $\Gamma_h=0.5$, $T_h=10$, $T_c=6$, $\omega_h=10$ and $\omega_c=7$. Solid gray, dashed blue and dotted red curves represent the case when $\lambda=0.1$, $\lambda=0.2$ and $\lambda=0.3$, respectively.}
\end{figure}

\subsection*{Effect of noise-induced coherence on efficiency at maximum power}
 
In Sec. III, we studied in detailthe effect of noise-induced coherence on the universal nature of efficiency. In Sec. III C, in strong-coupling and high-temperature regime, we  obtained analytic expressions for the EMP, which depends on the noise-induced coherence parameter $p$.  Here, we turn to study the effect of noise-induced coherence on the efficiency at maximum power of the heat engine. This information is also encoded in the second term of Eqs. (\ref{taylor1}), (\ref{taylor2}), and  (\ref{taylor3}), which is monotonically increasing function of $p$ for each case. This can also be verified directly by showing that derivative of $\eta^{\rm MP}_{\omega_h}$ (see Eq. (\ref{effwh})) ($\eta^*_{\omega_c}$ and $\eta^*_{\omega_h+\omega_c}$) with respect to $p$ is always positive. Hence, in the the strong-coupling and high-temperature regime, EMP of the engine increases with increasing noise-induce coherence parameter $p$. Hence, it is advisable to operate the engine at $p=1$ for achieving the maximum EMP in this operation regime. Our findings are illustrated through Fig. 3. In Fig. 3, we plot EMP as a function of $\eta_C$ for three different values of $p$, $p=-.9$, $p=0$, $p=.9$.
\begin{figure}   
 \begin{center}
\includegraphics[width=8.6cm]{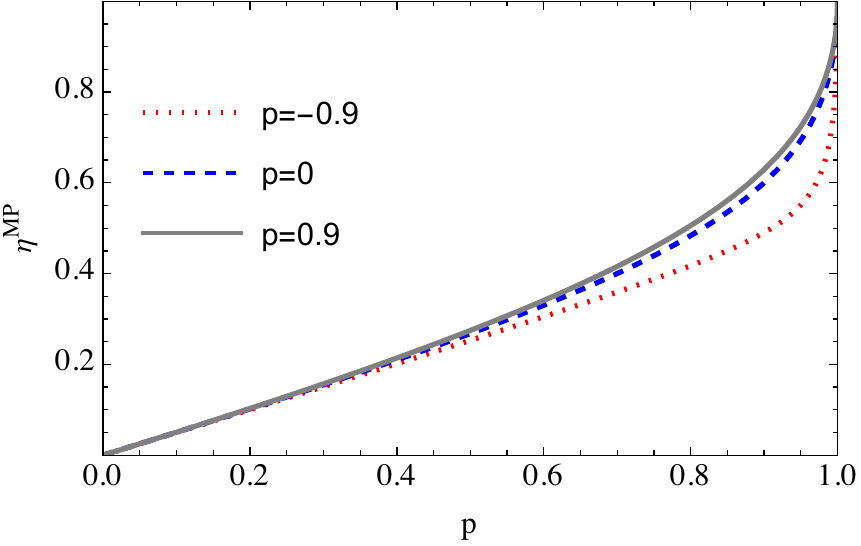}
 \end{center}
\caption{EMP (Eq. (\ref{effwh})) as a function of noise-induced coherence parameter $p$. Here, $\Gamma_c=1$, $\Gamma_h=0.5$, $T_h=10$, $T_c=6$, $\omega_h=10$ and $\omega_c=7$. }
\end{figure}

\section{Discussion and Conclusion}
 We started with two-parameter optimization of output power.  We observe that the presence of noise-induce coherence disrupts the left-right symmetry of the system, i.e.,  ($I(x,y) \neq I(x,y)$). Then, for both high-temperature and low-temperature regimes, we discussed the conditions  under which universal nature of efficiency can be retained.
  In conclusion, we have seen that the noise-induced coherence in a four-level heat engine influences both the universal characteristic of efficiency and its performance.   Furthermore, when we optimize the power with respect to one control parameter (keeping other one constant) in tight-coupling ($\lambda >> \Gamma_{h,c}$) limit in the high-temperature regime, we are unable to retain second universal term even when  the system satisfy flux reversal condition ($I(x, y)=-I(y, x)$). This difference can be attributed to the fact that the parametric space accessible to control frequencies of the system ($\omega_c,\omega_h$), is different for  two types of optimization schemes,   thus generating different outcomes. However, when we impose an additional symmetric constraint on the control frequencies, namely $\omega_c + \omega_h = k$ (or $\omega_c \omega_h = k$),  the universality of quadratic term $\eta_C^2/8$ can be established. We can conclude that in case of one parameter optimization, the left-right symmetry is not limited to system-bath coupling constants but should also include the permitted parametric space. In addition, we have shown that the strength of matter-field coupling plays an important role in quantifying the role of noise-induced coherence on the performance of a four-level heat engine. More precisely, the strength of matter-field coupling guides us to determine the suitable operational regime for the engine to maximize the benefits of noise-induced coherence. For instance, power increases monotonically with $p$ when the engine operates under strong-matter coupling ($\lambda\gg \Gamma_h$), in contrast to weak coupling ($\lambda\ll \Gamma_h$) where power decreases with $p$. Finally, we have shown that the EMP is monotonically increasing function of parameter $p$ when engine operates in high-temperature regime and strong-coupling regime. We conclude that strong coupling and high temperature regime is a favorable regime for the engine operation in the presence of noise-induced coherence in the system.
 
 \onecolumngrid

\appendix\section{Density matrix equations for four-level maser heat engine}
In the appendix, we present the density matrix equations  and their solution for the four-level maser heat engine. 
The total Hamiltonian of the four-level system interacting with a semi-classical field is given by \cite{Dorfman2018}
\begin{align}\label{eq:H01}
	H= H_0 + V(t)
\end{align}
where $H_0 = \hbar\sum_{i=g,0,1,2}\omega_i|i\rangle\langle i|,$ is the bare hamiltonian annd $V(t)=\hbar\lambda[e^{-i\omega t}(|1\rangle\langle 0|+|2\rangle\langle 0|)+e^{i\omega t}(|0\rangle\langle 1|+|0\rangle\langle 2|)]$ represents the semi-classical field applied between energy level $\vert 0\rangle$ and the degenrate levels $\vert 1\rangle$, $\vert 2\rangle$.
In a rotating frame, the system dynamics is described by the following  master equation:
\begin{equation}
\dot{\rho} = -\frac{i}{\hbar} [V_R,\rho] + \mathcal{L}_{h}[\rho] + \mathcal{L}_{c}[\rho],
\end{equation}
where $\mathcal{L}_{h(c)}$ represents the system-bath interaction with the hot (cold) reservoir and are known as dissipative Lindblad super-operators. The mathematical form of Lindblad super-operators is given by:
\begin{eqnarray}
\mathcal{L}_c[\rho]= \Gamma_c(n_c+1) \Big(A_c\rho A_c^\dagger  -\frac{1}{2}\big\{ A_c^\dagger A_k,\rho \big\}\Big) 
+  \Gamma_c n_c\Big(A_c^\dagger \rho A_c  -\frac{1}{2}\big\{ A_c A_c^\dagger,\rho \big\}\Big)  , \label{disscold}
\end{eqnarray}
\begin{eqnarray}
\mathcal{L}_h[\rho]=\sum_{k={1,2}}\Gamma_{hk}\Big[(n_h+1)\Big(A_k\rho A_k^\dagger  -\frac{1}{2}\big\{ A_k^\dagger A_k,\rho \big\}\Big) 
+  n_h\Big(A_k^\dagger \rho A_k  -\frac{1}{2}\big\{ A_k A_k^\dagger,\rho \big\}\Big)\Big] \nonumber
\\
+  \Gamma\cos\theta \Big[(n_h+1)\Big(A_1\rho A_2^\dagger -\frac{1}{2}\big\{ A_2^\dagger A_1,\rho \big\}\Big) 
 +
n_h\Big(A_1^\dagger \rho A_2  -\frac{1}{2}\big\{ A_2 A_1^\dagger,\rho \big\}\Big)\Big] \nonumber
\\
+  \Gamma\cos\theta \Big[(n_h+1)\Big(A_2\rho A_1^\dagger -\frac{1}{2}\big\{ A_1^\dagger A_2,\rho \big\}\Big) 
 +
n_h\Big(A_2^\dagger \rho A_1  -\frac{1}{2}\big\{ A_1 A_2^\dagger,\rho \big\}\Big)\Big]. \label{disshot}
\end{eqnarray} 
where  $A_c=\vert g\rangle \langle 0\vert$,  
$A_k=\vert g\rangle \langle k \vert$ ($k=1, 2$) are known as the jump operators and describes relevant transitions between energy levels.
The time evolution of the elements of density matrix equation is given by the following set of equations:
\begin{eqnarray}
	\dot{\rho}_{11}&=& i\lambda(\rho_{10}-\rho_{01})-\Gamma_{h}[(n_{h}+1)\rho_{11}-n_{h}\rho_{gg}] -\frac{1}{2} p \Gamma_h (n_h+1)(\rho_{12}+\rho_{21}),
\\
	\dot{\rho}_{22}&=& i\lambda(\rho_{20}-\rho_{02})-\Gamma_{h}[(n_h+1)\rho_{22}-n_h\rho_{gg}] - \frac{1}{2} p  \Gamma_h (n_h+1)(\rho_{12}+\rho_{21}),
\\
	\dot{\rho}_{00}&=& i\lambda(\rho_{01}+\rho_{02}-\rho_{10}-\rho_{20})\notag - \Gamma_c[(n_c+1)\rho_{00}-n_c\rho_{gg}],
\\
	\rho_{gg} &=& 1-\rho_{11}-\rho_{22}-\rho_{00},
\\
	\dot{\rho}_{12}&=&i\lambda(\rho_{10}-\rho_{02})-\frac{1}{2}[\Gamma_{h}(n_h+1)+\Gamma_{h}(n_h+1)]\rho_{12}
-\frac{1}{2} p \Gamma_h[(n_h+1)\rho_{11}
	 +(n_h+1)\rho_{22}-(n_h+n_h)\rho_{gg}], \nonumber
	 \\
\\
	\dot{\rho}_{10}&=&i\lambda(\rho_{11}-\rho_{00}+\rho_{12})-\frac{1}{2}[\Gamma_c(n_c+1)+\Gamma_h(n_h+1)]\rho_{10}	-\frac{1}{2} p \Gamma_h (n_h+1)\rho_{20},
\\
	\dot{\rho}_{20}&=&i\lambda(\rho_{22}-\rho_{00}+\rho_{21})-\frac{1}{2}[\Gamma_c(n_c+1)+\Gamma_{h}(n_h+1)]\rho_{20}
	-\frac{1}{2} p \Gamma_h(n_h+1)\rho_{10}.
\end{eqnarray}
In writing the above equations, we have dropped the subscript $R$ from $\rho$ for notational simplicity. 
In the steady state ($\dot{\rho}=0$), the above set of equations can be solved to yield
\begin{equation}
 \rho_{10}=\rho_{20}= -\frac{2 i \lambda  (p+1) \Gamma _c \Gamma _h \left(n_h+1\right) \left(n_c-n_h\right)}{A^\prime+B^\prime}, \quad \rho_{01}=\rho^*_{10},\label{solrho10}
\end{equation}
where 
\begin{eqnarray}
 A^\prime &=& (p+1) \Gamma _c \Gamma _h \left(n_h+1\right) \left(n_c \left(4 n_h+2\right)+3 n_h+1\right) \left(\Gamma _c \left(n_c+1\right)+(p+1) \Gamma _h \left(n_h+1\right)\right), \nonumber
 \\
  B^\prime &=& 8 \lambda ^2 \left(\Gamma _c \left(n_c \left(4 n_h+3\right)+2 n_h+1\right)+(p+1) \Gamma _h \left(4 n_h^2+5 n_h+1\right)\right).
\end{eqnarray}
Substituting Eq. (\ref{solrho10}) in Eq. (\ref{power222}), the expression for power can be derived as
\begin{equation}
    P= \frac{8\hbar \lambda^2 (p+1) \Gamma_h\Gamma_c \left(n_h+1\right) (n_h-n_c) (\omega_h-\omega_c)}{ A^\prime+ B^\prime}. \label{powerfinal}
\end{equation}

 \bibliography{SSD_NIC}

\bibliographystyle{apsrev4-2}

\end{document}